\documentstyle[11pt,epsfig,psfig]{article}
=1.60cm
\def\beq{\begin{equation}}
\def\eeq{\end{equation}}

\def\text#1{\mbox{\scriptsize #1}}

\makeatletter

\begin{document}

\title{RADIATION FROM A CHARGE CIRCULATING INSIDE A
WAVEGUIDE WITH DIELECTRIC FILLING}

\author{Anna S. Kotanjyan and Aram A. Saharian \footnote{E-mail: saharyan@www.physdep.r.am }  \\
{\it Institute of Applied Problems in Physics} \\
{\it 25 Nersessian Str., 375014 Yerevan, Armenia}}

\maketitle

\begin{abstract}
The emitted power of the radiation from a charged particle moving
uniformly on a circle inside a cylindrical waveguide is
considered. The expressions for the energy flux of the radiation
passing through the waveguide cross-section are derived for both
TE and TM waves. The results of the numerical evaluation are
presented for the number of emitted quanta depending on the
waveguide radius, the radius of the charge rotation orbit and
dielectric permittivity of the filling medium. These results are
compared with the corresponding quantities for the synchrotron
radiation in a homogeneous medium.
\end{abstract}

\section{Introduction}

The wide applications of the synchrotron radiation (see, for
example \cite{Sokolov,Ternov,Win80} and the references therein)
motivate the importance of investigations for various mechanisms
of control of the radiation parameters. From this point of view,
it is of interest to investigate the influence of a medium on the
spectral and angular distributions of the synchrotron emission.
This study is also important with respect to some astrophysical
problems \cite{Ryb79,Ahar00}. The radiation from a charged
particle circulating in a homogeneous medium was considered by
Tsytovich in \cite{Tsytovich} (see also \cite{Kitao,Zrelov}). It
had been shown that the interference between the synchrotron and
Cerenkov radiations leads to interesting effects. The radiation
from a charge rotating around a dielectric ball enclosed by a
homogeneous medium is investigated in papers \cite{Arzumanian,
Grigorian}. The presence of the ball leads to interesting effects:
if for substance of the ball and particle velocity the Cerenkov
condition is satisfied, there are strong narrow peaks in the
radiation intensity. At these peaks the radiated energy exceeds
the corresponding quantity in the case of a homogeneous medium by
some orders of magnitude. A similar problem for the case of the
cylindrical symmetry we have considered in Refs. \cite{grig,
Kotanjian}. In the paper \cite{grig} a recurrent scheme is
developed for constructing the Green function of the
electromagnetic field for a medium consisting of an arbitrary
number of coaxial cylindrical layers. The investigation of the
radiation from a charged particle circulating around a dielectric
cylinder immersed in a homogeneous medium \cite{Kotanjian}, has
shown that under the Cerenkov condition for the material of the
cylinder and the velocity of a charge there are narrow peaks in
the angular distribution of the number of quanta emitted into the
exterior space. For some values of parameters the density of the
number of quanta in these peaks exceeds the corresponding quantity
for the radiation in the vacuum by several orders.

The present paper is devoted to the radiation from a charge
rotating inside a cylindrical waveguide filled by a homogenous
dielectric. The contribution of waves of the electric and magnetic
type to the total radiation intensity is determined. The results
of the corresponding numerical calculations are presented and a
comparison with the intensity of the synchrotron radiation in a
homogenous medium is carried out.

\section{ Electromagnetic fields inside a waveguide}

To determine the electromagnetic field of a charged particle
rotating inside a cylindrical waveguide with conducting walls and
filled by a material with dielectric permittivity $\varepsilon
_{0}$, firstly we shall consider a more general problem. Let a
charge $q$ moves with the velocity $v$ along a circle of radius
$\rho _{0}$ in the plane $z=0$ inside a cylinder of radius $\rho
_{1}$ and axis $z$. The permittivity of the medium inside the
cylinder is $\varepsilon _{0}$ and it is immersed in a homogeneous
medium with permittivity $\varepsilon _{1}$ ( for simplicity we
shall assume that the permeability is equal to 1). The 4-potential
$A_{i}=(\varphi ,-\mathbf{A })$ of the corresponding
electromagnetic field can be determined via the Green function
(being a second rank tensor) by the formula
\begin{equation}
A_{i}(x)=-\frac{1}{2\pi ^{2}c}\int G_{il}(x,x^\prime
)j_{l}(x^\prime )d^{4}x^\prime ,\quad x=(t,{\mathbf{r}}),\quad
i,l=0,1,2,3, \label{vecpot}
\end{equation}
where a summation over index $l$ is assumed, $c$ is the speed of
light in the vacuum, and $j_{l}(x)$ is the 4-vector of the current
density of the charge. In the cylindrical coordinates $(\rho
,\varphi ,z)$ the spatial components of the latter are in form
\begin{equation}
j_{l}=\frac{vq}{\rho _{0}}\delta (\rho -\rho _{0})\delta (\varphi
-\omega _{0}t)\delta (z)\delta _{l2},\quad v=\omega _{0}\rho
_{0},\quad l=1,2,3 \label{hos}
\end{equation}
(the values of indices $l=1,2,3$ correspond to the coordinates
$\rho ,\varphi ,z$ respectively). A recurrent scheme for the
construction of the Green function in a medium consisting of an
arbitrary number of coaxial cylindrical layers is developed in
\cite{grig}. The Green function for the problem under
consideration can be derived from the general formulae of this
paper. By using the problem symmetry we can write the following
Fourier expansion for the Green function:
\begin{equation}\label{GreenFourier}
G_{il}(x,x')=\sum _{m=-\infty }^{+\infty }\int dk_zd\omega
G_{il}(\omega ,m,k_z,\rho ,\rho ')\exp \left\{ i\left[ m(\varphi
-\varphi ' )+k_z(z-z')-\omega (t-t')\right] \right\}.
\end{equation}
In the Lorentz gauge for the components we need here and for $\rho
=\rho _0<\rho _1 $, $\rho <\rho _1$ from the general formulae of
Ref. \cite{grig} one has
\begin{equation}\label{Greencomp}
G_{il}(\omega ,m,k_z,\rho ,\rho _0)=\frac{\pi
}{4i^{l-1}}\sum_{\alpha =-1,1}\alpha^{l}\left[ J_{m+\alpha
}(\lambda _0\rho _{<})H_{m+\alpha }(\lambda _0\rho _{>})
+B_{m}^{(\alpha )}J_{m+\alpha }(\lambda _0\rho )\right],\quad
l=1,2,
\end{equation}
where $\rho _{<}={\mathrm{min}}(\rho ,\rho _0)$, $\rho
_{>}={\mathrm{max}}(\rho ,\rho _0)$, $J_{m}(x)$ is the Bessel
function, $H_{m}(x)\equiv H_{m}^{(1)}(x)$ is the Hankel function
of the first kind. In (\ref{Greencomp}) the coefficients
$B_{m}^{(\alpha )}$ are defined by the expressions
\begin{eqnarray}
B_{m}^{(\alpha )} & = & -J_{m+\alpha }(\lambda _{0}\rho
_{0})\frac{W(H_{m+\alpha },H_{m+\alpha })}{W(J_{m+\alpha
},H_{m+\alpha })}+ \nonumber \\
 & & +\frac{i\alpha \lambda
_{1}H_{m}(\lambda _{1}\rho _{1})H_{m+\alpha }(\lambda _{1}\rho
_{1})}{\pi
\rho _{1}^{2}\beta _{1}W(J_{m+\alpha },H_{m+\alpha })}\sum_{p=\pm 1}\frac{%
J_{m+p}(\lambda _{0}\rho _{0})}{W(J_{m+p},H_{m+p})},
\label{bmalf}
\end{eqnarray}
where, as in the paper \cite{ Kotanjian}, we use notations
\begin{eqnarray}
\lambda _{1,0}&=&\frac{m\omega _{0}}{c}\sqrt{\varepsilon _{1,0}-\frac{%
c^{2}k_{z}^{2}}{m^{2}\omega _{0}^{2}}}  \label{lam} \\
\beta _{1}&=&\frac{\varepsilon _{0}}{\varepsilon _{1}-\varepsilon
_{0}}-\frac{\lambda _{0}}{2}J_{m}(\lambda _{0}\rho
_{1})\sum_{l=\pm 1}l\frac{H_{m+l}(\lambda _{1}\rho
_{1})}{W(J_{m+l},H_{m+l})}  \label{bet}
\end{eqnarray}
with
\begin{equation}
W(a,b)=a(\lambda _{0}\rho _{1})\frac{\partial b(\lambda _{1}\rho _{1})}{%
\partial \rho _{1}}-b(\lambda _{1}\rho _{1})\frac{\partial a(\lambda
_{0}\rho _{1})}{\partial \rho _{1}}.  \label{wronskian}
\end{equation}
In formula (\ref{furieobr}) the summand with $m=0$ does not depend
on time and hence does not give a contribution to the radiation
field. Therefore in the consideration of the radiation field we
can assume $m\neq 0$.

Substituting Eq. (\ref{hos}) and expressions (\ref{GreenFourier}),
(\ref{Greencomp}) for the Green function into formula
(\ref{vecpot}) we receive
\begin{equation}
A_{l}(x)=\sum_{m=-\infty }^{+\infty }e^{im(\varphi -\omega
_{0}t)}\int_{-\infty }^{\infty }dk_{z}e^{ik_{z}z}A_{ml}(k_{z},\rho
), \label{furieobr}
\end{equation}
where in the Lorentz gauge
\begin{eqnarray}
A_{ml}(k_{z},\rho ) &=&-\frac{v q}{4ci^{l-1}}\left\{
A_{ml}^{(0)}+\sum_{\alpha =\pm 1}\alpha ^{l}B_{m}^{(\alpha
)}J_{m+\alpha
}(\lambda _{0}\rho )\right\} ,\quad l=1,2,  \label{acomp} \\
A_{m3}(k_{z},\rho ) &=&0.  \nonumber
\end{eqnarray}
Here the term
\begin{equation}
A_{ml}^{(0)}=\sum_{\alpha =-1,1}\alpha ^{l}J_{m+\alpha }(\lambda
_{0}\rho _{<})H_{m+\alpha }(\lambda _{0}\rho _{>}),\label{potham}
\end{equation}
corresponds to the field of a charge in a homogeneous medium with
permittivity $\varepsilon _{0}$.

The vector potential of the electromagnetic field inside a
waveguide with conducting walls can be obtained from the formulae
given above taking the limit $\varepsilon _{1}\rightarrow \infty
$. From formulae (\ref{bmalf})-(\ref{wronskian}) it follows that
$\lambda _{1}(\varepsilon _{1})$ enters only in the argument of
the Hankel function. By a simple calculation in view of the
asymptotic formulae of this function for large values of the
argument \cite{Abramovic} it can be seen that in the limit
$\varepsilon _{1}\rightarrow \infty $ the coefficients
(\ref{bmalf}) take the form
\begin{eqnarray}
B_{m}^{(\alpha )}&=&-J_{m+\alpha }(\lambda _{0}\rho
_{0})\frac{H_{m+\alpha
}(\lambda _{0}\rho _{1})}{J_{m+\alpha }(\lambda _{0}\rho _{1})}- \nonumber \\
&& -\frac{%
i\alpha J_{m-\alpha }(\lambda _{0}\rho _{1})}{\pi \rho _{1}\lambda
_{0}J_{m}(\lambda _{0}\rho _{1})J^\prime _{m}(\lambda _{0}\rho
_{1})}\sum_{p=\pm 1}p\frac{J_{m+p}(\lambda _{0}\rho
_{0})}{J_{m+p}(\lambda _{0}\rho _{1})}. \label{bmlim}
\end{eqnarray}
Thus, the field inside a cylindrical waveguide is determined by formulae (%
\ref{furieobr}), (\ref{acomp}), where the coefficients
$B_{m}^{(\alpha )}$ are defined according to Eq. (\ref{bmlim}).
Note, that in the limit $\rho _{0}\rightarrow \rho _{1}$ one has
$B_{m}^{(\alpha )}=-H_{m+\alpha }(\lambda _{0}\rho _{0})$, and in
virtue of Eq. (\ref{potham}) fields (\ref{acomp}) tend to zero. We
could expect this result as when the charge approaches the wall of
a waveguide the particle and its mirror image cancel each other.
Before to proceed further calculations we shall consider the
analytical properties of $A_{ml}(k_{z},\rho )$ in Eq.
(\ref{furieobr}), as a function on the complex variable $k_{z}$.
From formulae (\ref{lam}), (\ref{potham}), (\ref{bmlim}) it can
seem, that the points $k_{z}=\pm m\omega _{0}\sqrt{\varepsilon
_{0}}/c$ are branchpoints for this function. However, by using the
expansions of the cylindrical functions (see, for example,
\cite{Abramovic}), it can be seen that actually $A_{ml}(k_{z},\rho
)$ is a function on $\lambda _{0}^{2}$ only:
$A_{ml}=A_{ml}(\lambda _{0}^{2},\rho )$. Note that the logarithmic
singularity $\ln \lambda _{0}$ contained in the expansion of the
Hankel function in $A_{ml}^{(0)}$ is cancelled with the
corresponding singularity of the first term in Eq. (\ref{bmlim}).
As a result the function $A_{ml}(k_{z},\rho )$ is meromorphic on
the complex plane $k_{z}$. From the formulae (\ref{potham}),
(\ref{bmlim}) it follows that this
function has poles corresponding to the zeros of the Bessel function $%
J_{m}(\lambda _{0}\rho _{1})$ and its derivative:
\begin{equation}
k_{z}=\pm k_{n}^{(\sigma )}\equiv \pm \sqrt{\frac{m^{2}\omega _{0}^{2}}{c^{2}%
}\varepsilon _{0}-\frac{j_{m,n}^{(\sigma )2}}{\rho _{1}^{2}},}\quad \sigma
=0,1,\quad J_{m}^{(\sigma )}(\lambda _{0}\rho _{1})=J_{m}^{(\sigma
)}(j_{m,n}^{(\sigma )})=0,\quad n=1,2,...,  \label{kazet}
\end{equation}
with the  notation $J_{m}^{(\sigma )}(x)=d^{\sigma
}J_{m}/dx^{\sigma }$. Note that the singularities in the first and
second terms on the right of formula (\ref{bmlim}) corresponding
to the zeros of the functions $J_{m\pm 1}(\lambda _{0}\rho _{1})$
are cancelled, and consequently the functions $B_{m}^{(\pm 1)}$
are regular at these points. In Eq. (\ref{kazet})
$j_{m,n}^{(\sigma )}$ are positive zeros of the Bessel function
($\sigma =0$) and its derivative ($\sigma =1$), arranged in
ascending order, $j_{m,n}^{(\sigma )}<j_{m,n+1}^{(\sigma )}$. All
these zeros are simple, and hence the values of $k_{z}$ given by
Eq. (\ref{kazet}) correspond to simple poles of the function
$A_{ml}(k_{z},\rho )$. They describe the eigenmodes of a
cylindrical waveguide and
are known as TM modes in the case $%
\sigma =0$ and TE modes in the case $\sigma =1$ \cite{Jekson}.

For real values of $\varepsilon _{0}$ poles (\ref{kazet}) are
situated on the real axis of the complex plane $k_{z}$, for
$j_{m,n}^{(\sigma )}\leq m\omega _{0}\rho _{1}\sqrt{\varepsilon
_{0}}/c$, and are purely imaginary otherwise. In formula
(\ref{furieobr}) to derive unambiguous result for the integral
over $k_{z}$ it is necessary to specify the rules to escape the
real poles (see Section \ref{sec:rad}). For this purpose note that
in physically realistic situations the permittivity is complex:
$\varepsilon _{0}=\varepsilon ^\prime_{0}+i\varepsilon ^{\prime
\prime }_{0}$, where the imaginary part $\varepsilon ^{\prime
\prime }_{0}>0$ describes the absorbtion in the medium. From here
it follows that the radicals $\pm k_{n}^{(\sigma )}$ from Eq.
(\ref{kazet}) with a positive/negative real part are situated in
the upper/lower half of the complex plane $k_{z}$. This leads to
the following rule to circle the poles in (\ref{furieobr}): in the
integral over $k_{z}$ the positive poles should be circled from
below, and negative ones -- from above. Note also that if the
walls of the waveguide have finite conductivity, then to the wave
number $k_{z}$ an additional imaginary part is added (see, for
example, \cite{Jekson}), which leads to the same rule to escape
the poles.

\section{ Radiation intensity}\label{sec:rad}

In this section we shall consider the radiation field travelling
inside the cylinder at large distances from the charge. First of
all let us show that in Eq. (\ref{acomp}) the fields
$A_{ml}^{(0)}$ and the term corresponding to the first summand in
definition (\ref{bmlim}) do not contribute to the radiation field.
This immediately follows from the estimate of the integral over
$k_{z}$ in Eq. (\ref{furieobr}) by using the stationary phase
method (see, for example, \cite{Fedoriuk}). As in this integral
the phase $k_{z}z$ has no stationary points, for large $\left|
z\right| $ the integral tends to zero faster than any power of
$1/\left| z\right| $, provided that the coefficient of the
exponential function belongs to the class $C^{\infty }(R)$. It
follows from here that the the singularities of this coefficient
can only give the contribution to the radiation field. As it has
been mentioned in the previous section, for the integral over
$k_{z}$ in Eq. (\ref{furieobr}) the singularities correspond to
the poles of the second summand in Eq. (\ref{bmlim}) at $k_{z}=\pm
k_{n}^{(\sigma )}$, defined by relations (\ref{kazet}). To
determine the corresponding contributions to the integral we note
that, as it has been mentioned above, by taking into account the
absorption, the points $k_{z}=k_{n}^{(\sigma )}(-k_{n}^{(\sigma
)})$ are situated in the upper (lower) half of the complex plane
$k_{z}$. Therefore the integration contour over $k_{z}$ in Eq.
(\ref{furieobr}) can be closed by a semicircle of a large radius
in the upper (lower) half-plane. This choice is stipulated by the
fact that for large $\left| z\right| $ the integrand exponentially
tends to zero in the upper (lower) half-plane for $z>0(<0)$. As a
result the corresponding integral vanishes when the radii of the
semicircles go to infinity. Thus, for large $\left| z\right| $,
$z>(<)0$ the integral over $k_{z}$ in (\ref{furieobr}), according
to the residue theorem, is equal to the sum of the residues at
poles $k_{z}=k_{n}^{(\sigma )}(-k_{n}^{(\sigma )})$ multiplied by
$2\pi i$. For large $n$, when $j_{m,n}^{(\sigma )}>m\omega
_{0}\rho _{1}\sqrt{\varepsilon _{0}}/c$,
the poles $\pm k_{n}^{(\sigma )}$ are purely imaginary in the limit $%
\varepsilon ^{\prime \prime }_{0}\rightarrow 0$ and the
corresponding contribution tends to zero exponentially for
$z\rightarrow \infty $. As a result these poles do not contribute
to the radiation field. Therefore, the radiation field far from
the charge takes the form
\begin{equation}
A_{l}({\mathbf{r}},t)=2\pi i\sum_{m=-\infty }^{+\infty
}e^{im(\varphi -\omega
_{0}t)}\sum_{\sigma =0,1}\sum_{n=1}^{n_{\max }^{(\sigma )}}{\mathrm{Res}}%
_{k_{z}=k_{n}^{(\sigma )}}A_{ml}e^{ik_{z}z},  \label{res}
\end{equation}
where the maximal value $n_{\max }^{(\sigma )}$ is defined by the
conditions
\begin{equation}
j_{m,n_{\max }^{(\sigma )}}^{(\sigma )}\leq \sqrt{\varepsilon _{0}}m\frac{v}{%
c}\frac{\rho _{1}}{\rho _{0}},\quad j_{m,n_{\max }^{(\sigma )}+1}^{(\sigma
)}>\sqrt{\varepsilon _{0}}m\frac{v}{c}\frac{\rho _{1}}{\rho _{0}}.
\label{jmax}
\end{equation}
Thus, at large distances from the charge the radiation field
inside the waveguide is presented in the form of waves with
discrete set of values of the wave vector projection  on the
waveguide axis, $k_{z}=k_{n}^{(\sigma )}$, $n=1,2,...,n_{\max
}^{(\sigma )}$, defined by formula (\ref{kazet}). Having $A_{ml}$
we can find the scalar potential by using the Lorentz gauge
condition. As a result the electric and magnetic fields can be
presented in the form of sums of the TE and TM waves:
\begin{equation}
F_{l}({\mathbf{r}},t)=\sum_{m=-\infty }^{+\infty }e^{im(\varphi
-\omega _{0}t)}\sum_{\sigma =0,1}\sum_{n=1}^{n_{\max }^{(\sigma
)}}F_{ml}^{(\sigma )}(j_{m,n}^{(\sigma )},\rho ),\quad F=E,H.
\label{fieldfurie}
\end{equation}
For the $z$-components of the fields corresponding to the
elementary waves one has
\begin{eqnarray}
E_{m3}^{(0)} &=&-\frac{2q}{\varepsilon _{0}\rho _{1}^{2}}\frac{%
J_{m}(j_{m,n}^{(0)}\rho _{0}/\rho _{1})}{J_{m}^{\prime
2}(j_{m,n}^{(0)})}J_{m}(j_{m,n}^{(0)}\rho /\rho _{1}),\quad
H_{m3}^{(0)}=0,\quad \mathrm{for\; TM\; waves},  \label{EH1} \\
H_{m3}^{(1)} &=&\frac{2qvi}{c\rho ^2
_{1}}\frac{j_{m,n}^{(1)2}J_{m}(j_{m,n}^{(1)}\rho /\rho
_{1})}{k_{n}^{(1)}(m^{2}-j_{m,n}^{(1)2})}\frac{J^\prime
_{m}(j_{m,n}^{(1)}\rho _{0}/\rho _{1})}{J_{m}^{2}(j_{m,n}^{(1)})}%
,\quad E_{m3}^{(1)}=0,\quad \mathrm{for\; TE\; waves}. \label{EH}
\end{eqnarray}
The transverse components can be found from the formulae (see, for
example, \cite{Jekson})
\begin{eqnarray}
{\mathbf{E}}_{mt}^{(0)} &=&\frac{ik_{n}^{(0)}\rho _{1}^{2}}{j_{m,n}^{(0)2}}%
{\mathbf{\nabla }}_{t}\Psi ,\quad
{\mathbf{H}}_{mt}^{(0)}=\frac{\varepsilon
_{0}m\omega _{0}}{ck_{n}^{(0)}}\left[ {\mathbf{e}}_{3}{\mathbf{E}}_{mt}^{(0)}%
\right] ,\quad  \nonumber \\
{\mathbf{H}}_{mt}^{(1)} &=&\frac{ik_{n}^{(1)}\rho _{1}^{2}}{j_{m,n}^{(1)2}}%
{\mathbf{\nabla }}_{t}\Psi ,\quad {\mathbf{E}}_{mt}^{(1)}=-\frac{m\omega _{0}}{%
ck_{n}^{(1)}}\left[ {\mathbf{e}}_{3}{\mathbf{H}}_{mt}^{(1)}\right]
, \label{EHgeks}
\end{eqnarray}
where $\Psi =E_{m3}^{(0)}$ for TM waves and $\Psi =H_{m3}^{(1)}$
for TE waves, ${\mathbf{\nabla }}_{t}=(\partial /\partial \rho
,im/\rho ,0)$, and $\mathbf{e}_3$ is the unit vector along the
axis $z$. The energy flux through the cross section of the
waveguide per unit time interval is given by the Poynting's vector
$\mathbf{S}$:
\begin{equation}
I=\int_{0}^{\rho _{1}}\rho d\rho \int_{0}^{2\pi }d\varphi ({\mathbf{e}}%
_{3}\cdot {\mathbf{S}}),\quad {\mathbf{S}}=\frac{c}{4\pi }\left[ {\mathbf{EH}}%
\right] ,  \label{pointing}
\end{equation}
where the fields are defined by expansions (\ref{fieldfurie}).
Substituting these expansions into Eq. (\ref{pointing}) and using
the formulae for the integrals involving products of the Bessel
functions [13] it can be seen that the contribution of the terms
$\left[ {\mathbf{E}}_{m}^{(\sigma )}(j_{m,n}^{(\sigma )},\rho
){\mathbf{H}}_{m' }^{(\sigma ' )^{\ast }}(j_{m' ,n'}^{(\sigma '
)},\rho )\right] $ is proportional to $\delta _{mm' }\delta _{nn'
}\delta _{\sigma \sigma '}$. As a result the intensity $I$ can be
presented as a sum of the radiation intensities on separate modes:
\begin{equation}
I=\sum_{m=1}^{\infty }\left( \sum_{n=1}^{n_{\max
}^{(1)}}I_{mn}^{\mathrm{(TE)}}+\sum_{n=1}^{n_{\max
}^{(0)}}I_{mn}^{\mathrm{(TM)}}\right) , \label{allint}
\end{equation}
where the energy radiated on the frequency $\omega =m\omega _{0}$
in the form of the TE and TM  modes per per unit time interval is
determined by the formulae
\begin{eqnarray}
I_{m}^{\mathrm{(TM)}} &=&\sum_{n=1}^{n_{\max
}^{(0)}}I_{mn}^{\mathrm{(TM)}},\quad
I_{mn}^{\mathrm{(TM)}}=\frac{2q^{2}vm}{\varepsilon _{0}\rho _{0}}\frac{k_{n}^{(0)}}{%
j_{m,n}^{(0)2}}\frac{J_{m}^{2}(j_{m,n}^{(0)}\rho _{0}/\rho _{1})}{%
J_{m+1}^{2}(j_{m,n}^{(0)})},  \label{intmod} \\
\quad I_{m}^{\mathrm{(TE)}} &=&\sum_{n=1}^{n_{\max
}^{(1)}}I_{mn}^{\mathrm{(TE)}},\quad
I_{mn}^{\mathrm{(TE)}}=\frac{2q^{2}v^{3}m}{c^{2}\rho _{1}^{2}\rho _{0}}\frac{%
j_{m,n}^{(1)2}}{k_{n}^{(1)}(j_{m,n}^{(1)2}-m^{2})}\frac{J
_{m}^{\prime 2}(j_{m,n}^{(1)}\rho _{0}/\rho
_{1})}{J_{m}^{2}(j_{m,n}^{(1)})}. \label{intmod1}
\end{eqnarray}
In these expressions the terms with fixed $n$ correspond to the
waves with a given value $k_{z}=$ $k_{n}^{(\sigma )}$ of the wave
vector projection on the waveguide axis. Note that in the limit
$\rho _{0}\rightarrow \rho _{1}$ the radiation goes to zero.
Expanding the Bessel functions in the numerators of Eqs.
(\ref{intmod}) and  (\ref{intmod1}) in this limit we receive
\begin{eqnarray}
I_{mn}^{{\mathrm{(TM)}}}& \approx &
\frac{2q^{2}\omega_{0}m}{\varepsilon
_{0}}k_{n}^{(0)}\left( 1-\frac{\rho _{0}}{\rho _{1}}\right) ^{2}, \label{as1}  \\
I_{mn}^{{\mathrm{(TE)}}} &\approx
&\frac{2q^{2}\omega_{0}^{3}m}{c^{2}}
\frac{(j_{m,n}^{(1)2}-m^{2})}{k_n^{(1)}}\left( 1-\frac{\rho
_{0}}{\rho _{1}}\right) ^{2} \label{as2}
\end{eqnarray}
under the assumption $j_{m,n}^{(\sigma )}(1-\rho _{0}/\rho
_{1})\ll 1$. As a necessary condition for the presence of a
radiation of a given type on the frequency $\omega =m\omega _{0}$
one has
\begin{equation}
n\omega _{0}>\omega _{m,1}^{(\sigma )},\quad \omega _{m,n}^{(\sigma )}=c%
\frac{j_{m,n}^{(\sigma )}}{\rho _{1}\sqrt{\varepsilon _{0}}},\label{omega}
\end{equation}
where we have introduced the boundary frequency $\omega
_{m,n}^{(\sigma )}$ for given $m,n$ and for a given type of waves
($\sigma =0,1$). In terms of the charge velocity it looks like
$v\sqrt{\varepsilon _{0}}/c>(j_{m,1}^{(\sigma )}/m)(\rho _{0}/\rho
_{1})$. In particular, by taking into account the inequality
$j_{m,1}^{(\sigma )}\geq m$ (see, for example, \cite{Abramovic}),
as a necessary condition for the presence of a radiation flux
through the cross-section of the waveguide we have $v\sqrt{
\varepsilon _{0}}/c>\rho _{0}/\rho _{1}$. If this condition does
not satisfied we have an interesting situation when the rotating
particle does not radiate. Note that the boundary frequency
$\omega _{m,n}^{(\sigma )}$ is a characteristic of the waveguide
and does not depend on the parameters of the charge (energy,
radius of the orbit). In formulae (\ref{intmod}) the quantities
$k_{n}^{(\sigma )}$ are expressed through the boundary frequency
by the formula
\begin{equation}
k_{n}^{(\sigma )}=\frac{\sqrt{\varepsilon _{0}}}{c}\sqrt{m^{2}\omega
_{0}^{2}-\omega _{m,n}^{(\sigma )2}}.  \label{kn}
\end{equation}
From Eqs. (\ref{intmod}), (\ref{kn}) it follows that if the charge
is orbiting with frequency $\omega _{0}=\omega _{m,n}^{(\sigma
)}/m$, i.e. if the frequency of radiation coincides with the one
of boundary frequencies (\ref {omega}), the radiation intensity
for the TE waves goes to infinity. Note that in this limit the
projection $k_{z}=k_{n}^{(\sigma )}$ goes to zero. However, it is
necessary to take into account that under these conditions the
absorption in the medium and in the walls of the waveguide becomes
important, and consequently it is necessary to take into account
the imaginary part $\varepsilon '' _{0}$ of the permittivity of
the filling medium and finite conductivity of the waveguide walls
(the damping constant in the waveguide walls depending on the
conductivity is presented, for example, in \cite{Jekson} ). Thus,
formulae (\ref{intmod}) are valid for frequencies, not too close
to the boundary ones, when $k_{n}^{(\sigma )}\gg \beta _{\lambda
},m\omega _{0} \sqrt{\varepsilon '' _{0}}/c$, where $\beta
_{\lambda }$ is the damping constant due to the ohmic losses in
the walls of the waveguide.

Introducing $\gamma_{m,n}^{(\sigma )}\equiv j_{m,n}^{(\sigma
)})\rho _{0}/m\rho _{1}$, the arguments of the Bessel functions in
the numerators of Eqs. (\ref{intmod}), (\ref{intmod1}) can be
written as $m\gamma _{m,n}^{(\sigma )}$. Note that for these
quantities one has the following inequalities  $\rho _{0}/\rho
_{1}<\gamma _{m,n}^{(\sigma )}\leq v\sqrt{\varepsilon _{0}}/c$.
Let us consider the radiation intensities (\ref{intmod}),
(\ref{intmod1}) for large values of $m$. From the well known
properties of the Bessel functions for large values of the order
it follows that the behavior of these intensities essentially
depends on the sign of $1-\gamma _{m,n}$.

First of all assume that the Cerenkov condition is not satisfied,
$v\sqrt{\varepsilon _{0}}/c<1$, and synchrotron radiation is
present only. For $1-v\sqrt{\varepsilon _{0}}/c\ll 1$ the
radiation intensity has maximum for $\omega \sim
\omega_{c}=m_{c}\omega_{0}$, with $m_{c}= (1-\gamma
_{m_c,n_{max}}^{(\sigma)2})^{-3/2}$, and as it follows from Eq.
(\ref{jmax}) $m_c<(1-v^2\varepsilon _0/c^2)^{-3/2}$. For the
frequencies $\omega \gg \omega _c$ the emitted power exponentially
decreases. If the Cerenkov condition is satisfied, for the modes
with $\gamma _{m,n}^{(\sigma )}>1$ in addition to the synchrotron
radiation we also have Cerenkov radiation. Note that in the
waveguide the sufficient condition for this radiation at a given
harmonic $m$ is $\gamma _{m,n_{max}}^{(\sigma )}>1$ instead of
$v\sqrt{\varepsilon _0}/c>1$. For this modes the radiation
intensity for large $m$ approximately linearly increases in the
mean with frequency. This is a characteristic feature of the
Cerenkov radiation. In this case to obtain a finite result for the
total intensity (\ref{allint}) it is necessary to take into
account the dispersion for the permittivity.
Consider now the limiting case of large values of the waveguide radius $%
\rho _{1}\rightarrow \infty $. In this limit the main contribution
to the radiation intensity is due to large values
$j_{m,n}^{(\sigma )}$, when we can use the asymptotic formula
\cite{Abramovic}
\begin{equation}
j_{m,n}^{(\sigma )}\sim \pi \left( n+\frac{m-1}{2}+\frac{(-1)^{\sigma }%
}{4}\right) .  \label{jmn}
\end{equation}
From this relation it follows that $\Delta \lambda
_{0}=(j_{m,n+1}^{(\sigma )}-j_{m,n}^{(\sigma )})/\rho _{1}=\pi
/\rho _{1}$. Now dividing and multiplying expressions
(\ref{intmod}) by $\pi /\rho _{1}$ and taking the limit $\rho
_{1}\rightarrow \infty $ the sum over $n$ can be replaced by the
integration over $\lambda _{0}$. As a result one has
\begin{eqnarray}
I_{0m}^{\mathrm{(TM)}}&=&\frac{q^{2}vm}{\varepsilon _{0}\rho _{0}}\int_{0}^{m\omega _{0}%
\sqrt{\varepsilon _{0}}/c}\frac{\sqrt{m^{2}\omega _{0}^{2}\varepsilon
_{0}/c^{2}-\lambda _{0}^{2}}}{\lambda _{0}}J_{m}^{2}(\lambda _{0}\rho
_{0})d\lambda _{0} \label{ihamTM},\\
I_{0m}^{\mathrm{(TE)}}&=&\frac{q^{2}v^{3}m}{c^{2}\rho _{0}}%
\int_{0}^{m\omega _{0}\sqrt{\varepsilon _{0}}/c}\frac{\lambda _{0}}{\sqrt{%
m^{2}\omega _{0}^{2}\varepsilon _{0}/c^{2}-\lambda _{0}^{2}}}
J_{m}^{\prime 2}(\lambda _{0}\rho _{0})d\lambda _{0}.
\label{iham}
\end{eqnarray}
By introducing a new integration variable $\theta $ according to
$\lambda _{0}\equiv $ ($m\omega _{0}\sqrt{\varepsilon _{0}}/c)\sin
\theta $ for the total radiation intensity (including the regions
$z>0$ and $z<0$) with the frequency $\omega =m\omega _{0}$ in the
limit $\rho _{1}\rightarrow \infty $ we receive
\begin{eqnarray}
I_{0m}&=&2(I_{0m}^{\mathrm{(TE)}}+I_{0m}^{\mathrm{(TM)}})
=\frac{2q^{2}m^{2}\omega _{0}^{2}}{c%
\sqrt{\varepsilon _{0}}}\times \nonumber \\
&& \times \int_{0}^{\pi /2}\left[ \cot^{2}\theta J_{m}^{2}(%
mv\sqrt{\varepsilon _{0}}\sin \theta /c)+\frac{v^{2}}{c^{2}}%
\varepsilon _{0}J_{m}^{\prime 2}(mv\sqrt{\varepsilon _{0}}\sin
\theta /c)\right] \sin \theta d\theta  \label{i0m}
\end{eqnarray}
This formula coincides with the expression for the radiation
intensity of a point charge circulating in a homogeneous medium
\cite{Tsytovich,Kitao}.

\begin{figure}[tbph]
\begin{center}
\epsfig{figure=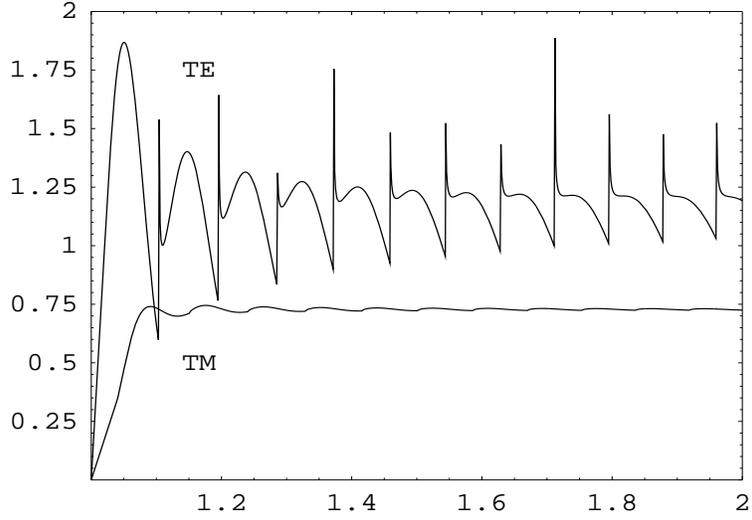,width=10cm,height=7cm}
\end{center}
\caption{ The number of quanta emitted on the harmonic $m=24$ in
the form of the TE and TM waves per circulating period of an
electron, multiplied by $\hbar c/q^2$, $\hbar
cN_m^{{\mathrm{(TE)}}}/q^2$ and $\hbar
cN_m^{{\mathrm{(TM)}}}/q^2$, versus the ratio $\rho _1/\rho _0$.
The energy of an electron is 2 MeV, and the dielectric
permittivity for the filling medium is $\varepsilon _0=3$.}
\label{fig1rad}
\end{figure}

\begin{figure}[tbph]
\begin{center}
\epsfig{figure=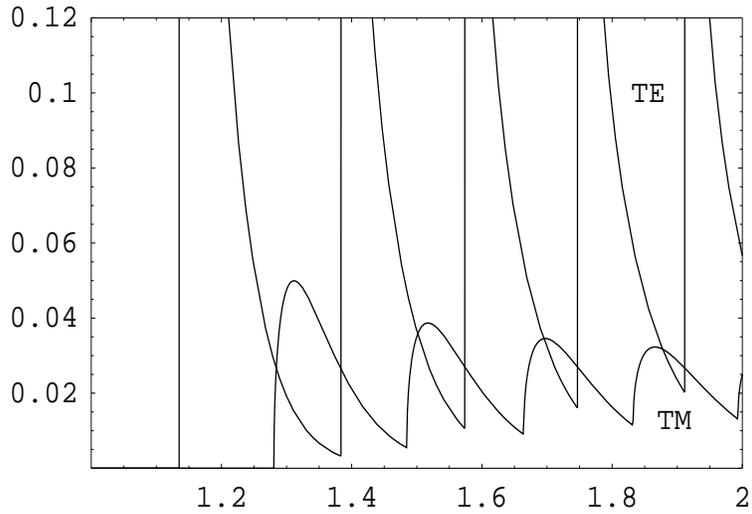,width=10cm,height=7cm}
\end{center}
\caption{ The same as in figure \ref{fig1rad} for $\varepsilon
_0=1 $.} \label{fig2rad}
\end{figure}

We carried out numerical calculations for the number of the
emitted quanta per one period of the particle orbiting,
\begin{equation}\label{quantanum}
N_{m}^{{\mathrm{(TE)}}}=\frac{2\pi }{\hbar m \omega
_0^2}I_{m}^{{\mathrm{(TE)}}},\quad
N_{m}^{{\mathrm{(TM)}}}=\frac{2\pi }{\hbar m \omega
_0^2}I_{m}^{{\mathrm{(TM)}}},
\end{equation}
for various values of the parameters $\varepsilon _{0}$, $\rho
_{1}/\rho _{0}$. In figures \ref{fig1rad} and \ref{fig2rad} the
quantities $N_{m}^{{\mathrm{(TE)}}}$, $N_{m}^{{\mathrm{(Tm)}}}$
for the harmonic $m=24$ are presented as functions on the ratio
$\rho _{1}/\rho _{0}$ for an electron with the energy $2MeV$ and
for the media with permittivities $\varepsilon _{0}=3$ and
$\varepsilon _{0}=1 $, respectively. The narrow peaks in the
graphs of the number of quanta for TE waves correspond to the
singularities on the boundary frequencies $\omega _{m,n}^{(\sigma
)}$. For the values of parameters corresponding to figure
\ref{fig1rad} one has $v\sqrt{\varepsilon _{0}}/c\approx 1.67$ and
the Cerenkov condition is satisfied.
As $j_{m,1}^{(0)}/m\approx 1.24$, $j_{m,1}^{(1)}/m\approx 1.1$,
then the condition (\ref{omega}) is satisfied for all values of
the ratio $\rho _1/\rho _0 \geq 1$. As a result the radiation is
nonzero for any $\rho _{1}/\rho _{0}>1$. For values of the
parameters corresponding to fig. 2 the Cerenkov's condition is not
satisfied ($v\sqrt{\varepsilon _{0}}/c\approx 0.97$) and the
radiation corresponds to the synchrotron emission.
Note that in this case the radiation is much more weaker in
comparison to the case of figure \ref{fig1rad}. Now there is a
range of values of the ratio $\rho _{1}/\rho _{0}$ for which
condition (\ref{omega}) is not satisfied and in this region the
radiation is absent. It is well seen in fig. \ref{fig2rad}. We
also carried out numerical calculations for an energy of an
election equal to 0.6 MeV and for the same values of the remaining
parameters, as in figure 1. In this case the Cerenkov condition is
not satisfied and, as calculations have shown, the qualitative
behaviour of the $N_m^{\mathrm{(TM)}}$ and $N_m^{\mathrm{(TE)}}$
in dependance of the ratio $\rho _{1}/\rho _{0}$ is similar to the
case presented in figure \ref{fig2rad}, and the radiation
intensity is much more weaker to compared with the case of figure
\ref{fig1rad}.

\begin{figure}[tbph]
\begin{center}
\begin{tabular}{ccc}
\epsfig{figure=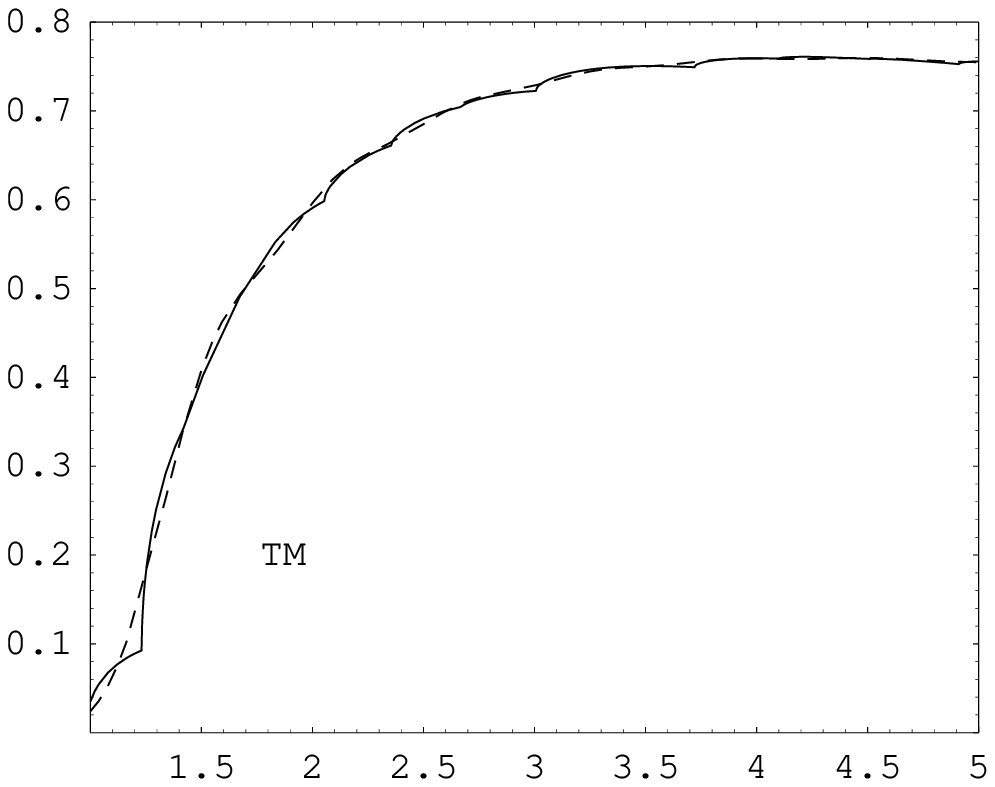,width=7cm,height=5cm} & \hspace*{0.5cm} & %
\epsfig{figure=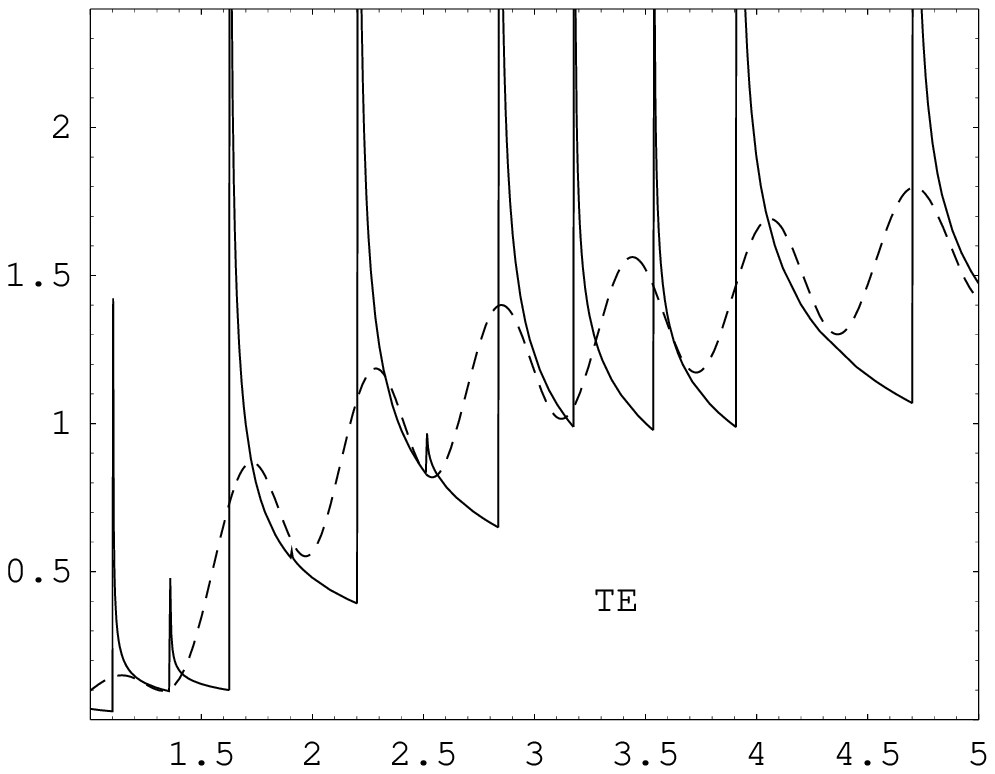,width=7cm,height=5cm}
\end{tabular}
\end{center}
\caption{The number of emitted quanta $\hbar cN_m/q^2$ for the TM
and TE waves on the harmonic $m=24$ as a function on the
dielectric permittivity $\varepsilon _0$ in the waveguide and in
the homogeneous medium (dashed lines). The energy of an electron
is equal to 2 MeV, and $\rho _1/\rho _0=1.5$.} \label{fig3rad}
\end{figure}

For comparison we have presented in figure \ref{fig3rad} the
dependances on the permittivity of the number of quanta emitted
from an electron in the form of the TM and TE waves in a waveguide
and in a homogeneous medium on the harmonic $m=24$. The energy of
an electron is $2$ $MeV$, and $\rho _{1}/\rho _{0}=1.5$.
Note that the intensities of the both types of radiation increase
when the permittivity increases.

\section*{Acknowledgments}
We are grateful to Professor A. R. Mkrtchyan, L. Sh. Grigoryan, S.
R. Arzumanian and H. F. Khachatryan for useful discussions and
suggestions. This work was supported in part by the Armenian
Ministry of Education and Science (Grant No. 1361).


\begin{thebibliography}{99}
\bibitem{Sokolov}  A. A. Sokolov and I. M. Ternov, {\it Relativistic electron}
(Nauka, Moscow, 1983) (in Russian).

\bibitem{Ternov}  I. M. Ternov, V. V. Mikhailin and V. R. Khalilov,
{\it Synchrotron Radiation and Its Applications} (Harwood Publ.,
New York, London, 1985).

\bibitem{Win80} {\it Synchrotron Radiation Research}, Ed. H.
Winick and S. Doniach (Plenum Press, New York, London, 1980).

\bibitem{Ryb79} G. B. Rybicky and A. P. Lightman, {\it Radiative Processes
in Astrophysics} (J. Wiley, New York, 1979).

\bibitem{Ahar00} F. A. Aharonian, {\it New Astron.}, {\bf 5}, 377
(2000).

\bibitem{Tsytovich}  V. N. Tsytovich, {\it Westnik MGU}, {\bf 11}, 27 (1951)(in Russian).

\bibitem{Kitao}  K. Kitao, {\it Progr. Theor. Phys.}, {\bf 23}, 759 (1960).

\bibitem{Zrelov}  V. P. Zrelov, {\it Wavilov--Cherenkov Radiation and Its
Applications in High-Energy Physics} (Atomizdat, Moscow, 1968)(in
Russian).

\bibitem{Arzumanian}  S. R. Arzumanian, L. Sh. Grigorian, Kh. V.
Kotanjian and A. A. Saharian, {\it Izv. Akad. Nauk Arm. SSR Fiz.},
{\bf 30}, 106 (1995) ({\it Sov. J. Contemp. Phys.}, {\bf 30},
No.3, 12 (1995)).

\bibitem{Grigorian}  L. Sh. Grigoryan, G. F. Khachatryan and S. R.
Arzumanyan, {\it Izv. Akad. Nauk Arm. SSR Fiz.}, {\bf 33}, 267
(1998) ({\it Sov. J. Contemp. Phys.}, {\bf 33}, No.6, 1 (1998)),
E-print:cond-mat/0001322.

\bibitem{grig}  L. Sh. Grigorian, A. S. Kotanjian and A. A. Saharian,
{\it Izv. Akad. Nauk Arm. SSR Fiz.}, {\bf 30}, 239 (1995) ({\it
Sov. J. Contemp. Phys.}, {\bf 30}, No.6, 1 (1995)).

\bibitem{Kotanjian}  A. S. Kotanjyan, H. F. Khachatryan, A. V. Petrosyan
and A. A. Saharian, {\it Izv. Akad. Nauk Arm. SSR Fiz.}, {\bf 35},
115 (2000) ({\it Sov. J. Contemp. Phys.}, {\bf 35}, No.3, 1
(2000)).

\bibitem{Abramovic} {\it Handbook of Mathematical Functions}, edited by
M. Abramowitz and I. A. Stegun (Dover, New York, 1972).

\bibitem{Jekson} J. D. Jackson, {\it Classical Electrodynamics }
(John Wiley and Sons, 1962).

\bibitem{Fedoriuk} M. V. Fedoryuk, {\it Asymptotics: Integrals and
Series} (Nauka, Moscow (1987)) (in Russian).

\bibitem{Prudnikov} A. P. Prudnikov, Yu. A. Brychkov and O. I.
Marichev, {\it Integrals and Series} (Gordon and Breach, New York,
1986), Vol.2.

\end{thebibliography}
\end{document}